\documentclass[conference]{IEEEtran}
\usepackage{graphicx}
\usepackage{amssymb, amsmath,amsthm, amsfonts}
\usepackage{tikz}
\usepackage{verbatim}
\usepackage{pifont}

\newtheorem{theorem}{Theorem}%[section]

\newtheorem{lemma}[theorem]{Lemma}

\newtheorem{defn}{Definition}

\newcommand{\be}{\begin{equation}}
\newcommand{\ee}{\end{equation}}
\newcommand{\ben}{\begin{equation*}}
\newcommand{\een}{\end{equation*}}
\newcommand{\ba}{\begin{eqnarray}}
\newcommand{\ea}{\end{eqnarray}}

\newcommand{\suppress}[1]{}

\def\fb{\sf fb} %feedback symbol
\def\sbc{\sf sbc} %scalar BC symbol
\def\vbc{\sf vbc} %vector BC symbol
\def\noisyfb{\sf noisy-fb} %Noisy FB symbol
\def\pfb{\sf partial-fb} %Noisy FB symbol
\def\nofb{\sf wo-fb} %No FB symbol

\def\h2{\tilde h}
\newcommand{\sw}{\sigma_{2}^2} %noise variance of Z_2
\renewcommand{\ss}{\sigma_{1}^2} %noise variance of Z_1
\newcommand{\vs}{\varsigma^2} %noise variance of Z
\renewcommand{\vss}{\varsigma_{1}^2} %noise variance of \tilde{Z}_1
\newcommand{\vsw}{\varsigma_{2}^2} %noise variance of \tilde{Z}_2
\def\sas{\sigma_{11}^2} %noise variance of antenna a
\def\sbs{\sigma_{12}^2} % noise variance of antenna b
\def\se{\sigma_e^2} % noise variance of antenna b
\def\sfb{\sigma_{\fb}^2} % noise variance of antenna b
\def\eE{\mathbb E} % noise variance of antenna b

\DeclareMathOperator{\Var}{Var}

\title{On the Noisy Feedback Capacity of Gaussian Broadcast Channels}

\author{
\IEEEauthorblockN{Sibi Raj~B.~Pillai}
\IEEEauthorblockA{Department of Electrical Engineering\\
	Indian Institute of Technology Bombay\\
         bsraj@ee.iitb.ac.in }
\and
\IEEEauthorblockN{Vinod~M.~Prabhakaran}
\IEEEauthorblockA{School of Technology and Computer Science\\
	Tata Institute of Fundamental Research, Mumbai\\
	vinodmp@tifr.res.in}
}

\begin{document}

\maketitle 

\begin{abstract}

It is well known that, in general, feedback may enlarge the capacity region
of Gaussian broadcast channels. This has been demonstrated even when the
feedback is noisy (or partial-but-perfect) and only from one of the
receivers. The only case known where feedback has been shown not to enlarge
the capacity region is when the channel is physically degraded. In this paper, we
show that for a class of two-user Gaussian broadcast channels (not
necessarily physically degraded), passively feeding back the stronger
user's signal over a link corrupted by Gaussian noise does not enlarge the
capacity region if the variance of feedback noise is above a certain
threshold.
\end{abstract}

\section{Introduction}

It is well known that  feedback does not increase the capacity of a memoryless
point-to-point channel, a result which goes back to C.~E.~Shannon
\cite{Shannon56}. However, feedback has a positive impact in simplifying coding
schemes and boosting error exponents \cite{GamalKim11}. With the discovery of
capacity regions for several multiuser models in the '70s and '80s, it was of
interest to find the impact of feedback in these models. For the discrete
memoryless broadcast channel (BC), El~Gamal~\cite{Gamal78} showed that feedback
does not enlarge the capacity region when the channel is physically degraded,
and Dueck~\cite{Dueck80} demonstrated a broadcast channel for which points
outside its capacity region can be attained using feedback.

For a two-user scalar Gaussian broadcast channel (GBC),
El~Gamal~\cite{Gamal81b} showed that the capacity region is unchanged by the
presence of noiseless feedback, if one of the receivers is physically degraded
with respect to the other.  However, Ozarow and Leung~\cite{Ozarow84b} showed
the surprising fact that the GBC capacity region is enlarged by feedback for a
class of positively correlated noise processes at the receivers.  The technique
of~\cite{Ozarow84b} used full causal feedback from both the receivers.
Recently it was shown that even perfect causal feedback from one of the
receivers can enlarge the GBC capacity region~\cite{Sibi08}. In other related
works, it was shown that the capacity enlargement can occur for the discrete
memoryless broadcast channel even when the feedback is
noisy~\cite{Shayewitz13},~\cite{Pradhan13}, or
rate-limited~\cite{Wigger13,Wigger14}. The GBC case was also considered
in~\cite{Pradhan13} and~\cite{Wigger14}. For the Gaussian multiple-access channel
(MAC), it is known that noisy feedback, even to only one of the transmitters,
always enlarges the capacity region~\cite{LapWig10}. Moreover, a duality has
been shown between linear coding schemes for MAC and BC in the presence of
noiseless feedback~\cite{AmStWi14}. Thus, in general, there is an optimism
about the availability of feedback enlarging the capacity region of broadcast
channels when the channel is not physically degraded. 

Against this backdrop, the purpose of this paper is to
show that the anticipated capacity enlargement may not exist for 
all feedback models over a GBC.
We show that, for a class of two-user scalar GBCs, when the stronger receiver's
signal (i.e., the signal of the receiver with the smaller noise variance) is
passively fed back to the transmitter over a noisy link corrupted by
independent Gaussian noise, any capacity enlargement is impossible if the
feedback noise variance is above a certain threshold. Our class of channels is,
in fact, the same as that studied by Ozarow and Leung~\cite{Ozarow84b} where
they showed that the capacity region is enlarged by perfect feedback from both
receivers. This class includes independent noises at the two users as well as a
range of positive correlations.

We also study a related class of vector GBCs with partial-but-perfect feedback.
Specifically, consider a vector GBC with a strong receiver employing two
receive antennas and the other receiver and encoder only having single
antennas.  In the absence of feedback, the optimal scheme is super-position
coding, along with maximal ratio combining at the strong receiver. Now, assume
that the weak user has a higher noise variance with respect to the  first
antenna of the strong receiver. Then, we will show that even perfectly feeding
back the second antenna output from the strong receiver does not enlarge the
capacity region.

The organization of this paper is as follows. In the next section, we will
describe the broadcast channel and the feedback model. This will be followed by
our main result in Section~\ref{sec:thresh}, which shows that feedback from the
stronger user does not enlarge the capacity region if the feedback noise
variance is above a certain threshold for a class of Gaussian noise models. A
Gaussian BC with two antennas at the strong receiver and perfect feedback from
one of the antennas will be considered in Section~\ref{sec:mult}.
Section~\ref{sec:conc} concludes the paper.

\section{System Model} \label{sec:model}

Consider a memoryless two-user scalar Gaussian broadcast channel, where
receiver~$1$ is the stronger receiver (i.e., its noise variance is not larger
than the noise variance of the other user). Assume a noisy feedback link from
the strong receiver to the transmitter, as shown in Figure~\ref{fig:bc:one}. 
\usetikzlibrary{arrows}
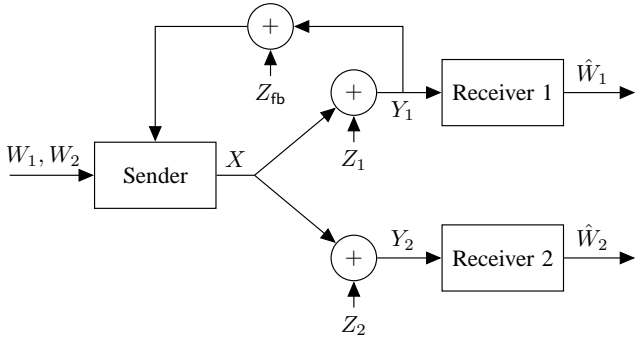
\begin{figure}[htbp]
\begin{center}
\scalebox{0.88}{
\begin{tikzpicture}[line width=0.5pt, >=triangle 45, text centered]
\coordinate (cenc) at (-0.5,0);
\coordinate  (cdec1) at  (4.75,1.25);
\coordinate  (cdec2) at  (4.75,-1.25);
\coordinate  (cz1) at  (2.5,1.25);
\coordinate  (cz2) at  (2.5,-1.25);

\node (enc) at (cenc)[rectangle, draw, text width=1.6cm, minimum height=1cm]{Sender};
\node (z1)  at (cz1)[circle, draw] {$+$};
\node (z2)  at (cz2)[circle, draw] {$+$};
\node (dec1) at (cdec1)[rectangle, draw, text width=1.6cm, minimum height=1.0cm]{Receiver~1};
\node (dec2) at (cdec2)[rectangle, draw, text width=1.6cm, minimum height=1.0cm]{Receiver~2};
\draw[->] (z1) -- node (cfb) [pos=0.4]{} (dec1); \node at (cfb)[below]{$Y_1$};
\draw[->] (z2) --node[above, pos=0.4]{$Y_{2}$} (dec2);
\draw[<-] (enc) -- node[above, pos=0.6]{$W_1, W_2$} ++(-2.2,0) ;
\draw[->] (enc) -- node[above, pos=0.5]{$X$} ++(1.5,0) -- (z1);
\draw[->] (enc)++(1.5,0) -- (z2);
\path (cfb) --++(0,1) --++(-2,0) node (zfb)[circle, draw]{$+$};
\draw[->] (cfb.center) |- (zfb);
\draw[->] (zfb) -| (enc);
\draw[->] (dec1) -- node[above, pos=0.4]{$\hat W_1$} ++(2,0);
\draw[->] (dec2) -- node[above, pos=0.4]{$\hat W_2$} ++(2,0);
\draw[<-] (z1) --++(0,-0.75)node[below]{$Z_1$};
\draw[<-] (z2) --++(0,-0.75)node[below]{$Z_2$};
\draw[<-] (zfb) --++(0,-0.75)node[below]{$Z_{\fb}$};
\end{tikzpicture}
}
\caption{Scalar Gaussian broadcast channel  with noisy feedback (causal) \label{fig:bc:one}}
\end{center}
\end{figure}

In this model, $X$ represents the transmitted signal. The additive
(forward) channel noises $Z_1, Z_2$ are zero-mean jointly Gaussian with
variances $\ss, \sw$, respectively, and correlation
coefficient $\rho$. As mentioned above, we will take $\ss \leq \sw$. The
additive noise on the feedback channel $Z_{\fb}$ is assumed to be
independent of $(Z_1,Z_2)$ and zero-mean Gaussian with variance $\sfb$.
In this setup, we aim to send two independent messages, say $W_1$ and
$W_2$, to the respective receivers.  The transmitted symbol at the
$i^{th}$ instant can be a function of the messages and the causal but noisy
feedback of the stronger user's signal, i.e.,
\begin{align} \label{eq:in:symb}
X_i = g_i(W_1, W_2, Y_{1}^{i-1}+Z_{\fb}^{i-1}),
\end{align} 
where the symbol $U^{i-1}  \triangleq (U_1, \cdots, U_{i-1})$. We will convey
a pair of messages in $n$ uses of the channel, and the alphabet sizes of
$W_1,W_2$ are $2^{nR_1}$ and $2^{nR_2}$, respectively. Consider the average
probability of error over a uniform choice of the messages, which we denote
by $P_e(n)$. We are interested in the capacity region $\mathcal
C_{\sbc}^{\noisyfb}$ of our model under an average transmit power constraint of
$P$.
\begin{defn}
The capacity region $\mathcal C_{\sbc}^{\noisyfb}$ is the closure of the set of
all rate-pairs $(R_1,R_2)$
such that there is a sequence of encoder-decoders with $P_e(n)
\rightarrow 0$ as $n\rightarrow \infty$.
\end{defn}
Let $\mathcal C_{\sbc}^{\nofb}$ denote the capacity region without
feedback. Clearly,
$$
\mathcal C_{\sbc}^{\nofb} \subseteq \mathcal C_{\sbc}^{\noisyfb}.
$$
The region $\mathcal C_{\sbc}^{\nofb}$ is a well known quantity, which can be achieved
by superposition coding~\cite{CovTho91}. 

\section{Feedback Noise Variance Thresholds} \label{sec:thresh}
Our main result establishes a threshold for the feedback noise variance
beyond which the feedback from  stronger user does not enlarge the
capacity region for a class of forward noise correlations (including
independent and physically degraded cases).
\begin{theorem} \label{thm:fb:gbc}
For $0 \leq \rho \leq \sqrt{\frac{\ss}{\sw}} \leq 1$,
$\mathcal C_{\sbc}^{\noisyfb} = \mathcal C_{\sbc}^{\nofb}$ whenever
\begin{align} \label{eq:fb:gbc}
\frac{\sfb}{\ss} \geq \frac{1}{\frac{\sw}{\ss} -1}\left( 1 - \rho\sqrt{\frac{\sw}{\ss}}\right)^2.
\end{align}
\end{theorem}

\begin{figure}
\begin{center}
\usetikzlibrary{decorations.markings}
\begin{tikzpicture}[label/.style={%
   postaction={ decorate,transform shape,
   decoration={ markings, mark=at position .5 with \node #1;}}}]
    \fill[fill=gray] (0,0) -- (0,5.2) -- plot [domain={(sqrt(1+4*5.2)-1)/2}:0] (\x,{\x+\x^2})  -- cycle;
    \draw plot[domain=0:{(sqrt(1+4*5.2)-1)/2}] (\x,{\x+\x^2});
    \draw[label={[above]{$\ss=\sw$}},thin] plot[domain=0:5.2] (\x,\x);
        \draw [thin] (0,0) --  coordinate (xaxis) (5.2,0) -- (5.2,5.2) --
(0,5.2) -- coordinate (yaxis) (0,0);
    	\foreach \x in {0,2,4}
     		\draw (\x,1pt) -- (\x,-3pt)
			node[anchor=north] {\x};
    	\foreach \y in {0,2,4}
     		\draw (1pt,\y) -- (-3pt,\y) 
     			node[anchor=east] {\y}; 
	\node[below=0.5cm] at (xaxis) {${\ss}/{\sfb}$};
	\node[rotate=90,above=0.5cm] at (yaxis) {${\sw}/{\sfb}$};
\end{tikzpicture}
\end{center} 
\caption{Theorem~\ref{thm:fb:gbc} guarantees that for independent forward
channel noises, noisy feedback from the stronger user does not enlarge the
capacity region when the noise variances fall in the shaded region.}
\label{fig:sbc-plot}
\end{figure}
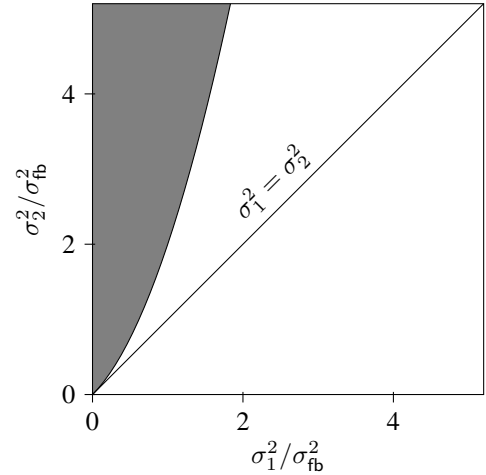

Notice that, when $\rho=\sqrt{\frac{\ss}{\sw}}$, the broadcast channel is
physically degraded and the feedback noise variance threshold in
\eqref{eq:fb:gbc} is 0. This is already implied by El~Gamal's
result~\cite{Gamal81b}. At the other extreme, when the forward channel noises
are independent $(\rho=0)$, the result is as shown in Fig.~\ref{fig:sbc-plot}.
However, note that the theorem does not give a threshold for all positively
correlated $Z_1,Z_2$. For example, when $Z_2=2Z_1$, we have $1=\rho \ge
\sqrt{\frac{\ss}{\sw}}=\frac{1}{2}$, and Theorem~\ref{thm:fb:gbc} does not apply. More
specifically, the theorem above gives a threshold whenever we can write
$Z_1,Z_2$ as \eqref{eq:equiv1}-\eqref{eq:equiv2}, where
$Z,\tilde{Z}_1,\tilde{Z}_2$ are independent. This is also the class of channels
studied by Ozarow and Leung~\cite{Ozarow84b}.

\begin{IEEEproof}
We start by noting that when $\ss \leq \sw$ and $\rho \leq
\sqrt{\frac{\ss}{\sw}}$,  without loss of generality, we may write
$Z_1,Z_2$ in the form
\begin{align}
Z_1 &= Z + \tilde{Z}_1 \label{eq:equiv1}\\
Z_2 &= Z + \tilde{Z}_2,\label{eq:equiv2}
\end{align}
where $Z,\tilde{Z}_1,\tilde{Z}_2$ are independent zero-mean Gaussian random variables
with variances $\vs,\vss,\vsw$, respectively,
given by
\begin{align*}
\vs &= \rho\sqrt{\ss\sw}\\
\vss &= \ss - \rho\sqrt{\ss\sw}\\
\vsw &= \sw - \rho\sqrt{\ss\sw}.
\end{align*}

The key idea of the proof is to give receiver~1 access to the feedback
noise (step (a) below). By Fano's inequality, (suppressing an $n\epsilon$ term)
\begin{align*}
nR_1 &\approx I(W_1;Y_1^n|W_2)\\
 &\stackrel{\text{(a)}}{\leq} I(W_1;Y_1^n,Z_{\fb}^n|W_2)\\
 &= \sum_{i=1}^n I(W_1;Y_{1i}|W_2,Y_1^{i-1},Z_{\fb}^{i-1})\\
 &= \sum_{i=1}^n\!\! h(Y_{1i}|W_2,Y_1^{i-1}\!\!\!,Z_{\fb}^{i-1}) -
h(Y_{1i}|W_1,W_2,Y_1^{i-1}\!\!\!,Z_{\fb}^{i-1})\\
 &\stackrel{\text{(b)}}{\leq} 
 \sum_{i=1}^n h(Y_{1i}|W_2,Y_1^{i-1} - \alpha Z_{\fb}^{i-1}) - h(Z_{1i})\\
 &= \sum_{i=1}^n h(Y_{1i}|W_2,X^{i-1}\!\!+Z^{i-1}\!\!+\tilde{Z}_1^{i-1}\!\!-\alpha Z_{\fb}^{i-1}) %\notag \\
- h(Z_{1i}),
\end{align*}
where in (b) we used the fact that $X_i$ is a function of $
(W_1,W_2,Y_1^{i-1},Z_{\fb}^{i-1})$. We choose $\alpha=\frac{\vss}{\sfb}$ so
that $\tilde{Z}_1-\alpha Z_{\fb}$ is independent of $\tilde{Z}_1+Z_{\fb}$.
This choice ensures that $\tilde{Z}_1-\alpha Z_{\fb}$ is independent of
$Z_1+Z_{\fb}=Z+\tilde{Z}_1+Z_{\fb}$. Notice that (i) $X_i$ is a function only of
everything the encoder knows before time instant~$i$, namely,
$(W_1,W_2,Z_1^{i-1}+Z_{\fb}^{i-1})$, and (ii) both
$\tilde{Z}_1^{i-1}-\alpha Z_{\fb}^{i-1}$ and $\tilde{Z}_2^{i-1}$ are
independent of encoder's knowledge $(W_1,W_2,Z_1^{i-1}+Z_{\fb}^{i-1})$
before time instant~$i$. Now, if  $\Var(\tilde{Z}_1-\alpha Z_{\fb}) \leq
\Var(\tilde{Z}_2)$, for each $i$, then we can write% 
\footnote{\label{fn:fb:sbc}To see this,
let us write $\tilde{Z}_2=\tilde{Z}_{2a}+\tilde{Z}_{2b}$, where
$\tilde{Z}_{2a},\tilde{Z}_{2b}$ are independent and also independent of
everything else. Let $\Var(\tilde{Z}_{2a})=\Var(\tilde{Z}_1-\alpha
Z_{\fb})$. Then, we may write
\begin{multline*}
h(Y_{1i}|W_2,X^{i-1}+Z^{i-1}+\tilde{Z}_1^{i-1}-\alpha Z_{\fb}^{i-1})  \\ \leq
h(Y_{1i}|W_2,X^{i-1}+Z^{i-1}+\tilde{Z}_{2a}^{i-1}),
\end{multline*}
since the independence of $(\tilde{Z}_1^{i-1}-\alpha Z_{\fb}^{i-1})$ and
everything the encoder has access to before time instant~$i$ allows us
to replace $\tilde{Z}_1^{i-1}-\alpha Z_{\fb}^{i-1}$ by any other
memoryless Gaussian random vector ($\tilde{Z}_{2a}^{i-1}$ in our case) with the same
variance and which is also independent of encoder's knowledge before
time~$i$. Now, using the fact that $Z_{2b}^{i-1}$ is independent of
everything else, we have
\begin{align*}
&h(Y_{1i}|W_2,X^{i-1}+Z^{i-1}+\tilde{Z}_{2a}^{i-1})\\
&\hspace{3cm}=
h(Y_{1i}|W_2,X^{i-1}+Z^{i-1}+\tilde{Z}_{2a}^{i-1},\tilde{Z}_{2b}^{i-1})\\
&\hspace{3cm}\leq  
h(Y_{1i}|W_2,X^{i-1}+Z^{i-1}+\tilde{Z}_{2a}^{i-1}+\tilde{Z}_{2b}^{i-1}).
\end{align*}
}
\begin{align*} 
& h(Y_{1i}|W_2,X^{i-1}+Z^{i-1}+\tilde{Z}_1^{i-1}-\alpha Z_{\fb}^{i-1})\\
&\hspace{3cm}
\leq h(Y_{1i}|W_2,X^{i-1}+Z^{i-1}+\tilde{Z}_2^{i-1}).
\end{align*}
It is easy to verify that the condition $\Var(\tilde{Z}_1-\alpha
Z_{\fb}) \leq \Var(\tilde{Z}_2)$ is precisely \eqref{eq:fb:gbc}.
Substituting back, we have
\begin{align}
nR_1 \leq \left(\sum_{i=1}^n h(Y_{1i}|W_2,Y_2^{i-1})\right) - \frac{n}{2}\log 2\pi e\ss. \label{eq:sbc:R1}
\end{align}

By Fano's inequality,
\begin{align}
nR_2 &\approx I(W_2;Y_2^n)\notag\\
 &= h(Y_2^n) - h(Y_2^n|W_2)\notag\\
 &\leq  \sum_{i=1}^n h(Y_{2i}) - h(Y_2^n|W_2)\notag\\
 &\leq \frac{n}{2}\log 2\pi e(P+\sw) - h(Y_2^n|W_2)\label{eq:sbc:R2}.
\end{align}
where the last inequality follows (via concavity of $\log$) from the power
constraint and the memorylessness of the channel. We can relate the first
term of \eqref{eq:sbc:R1} and the second term above through the entropy
power inequality (EPI).
\begin{lemma}\label{lem:epi}
\begin{align}
2^{\frac{2}{n}h(Y_2^n|W_2)} \geq 2^{\frac{2}{n}\sum_{i=1}^n
h(Y_{1i}|W_2,Y_2^{i-1})} + 2\pi e (\sw-\ss). \label{eq:sbc:epi}
\end{align}
\end{lemma}
The proof, which is along the lines of El~Gamal~\cite[Lemma 1]{Gamal81b} and
Blachman~\cite{Blachman65}, is given in the appendix.

To finish the proof, we note that
\[ \frac{n}{2}\log2\pi e(P+\sw) \geq h(Y_2^n) \geq h(Y_2^n|W_2) \geq
h(Z_2^n),\]
 where the last inequality follows from $h(Y_2^n|W_2) = \sum_{i=1}^n
h(Y_{2i}|W_2,Y_2^{i-1}) \leq \sum_{i=1}^n h(Y_{2i}|X_i,W_2,Y_2^{i-1}) =
h(Z_2^n)$. Hence, there is some $\theta\in[0,1]$ such that
\[ \frac{1}{n}h(Y_2^n|W_2) = \frac{1}{2}\log 2 \pi e (\theta P + \sw).\]
Substituting in \eqref{eq:sbc:R2}, we get
\begin{align*}
R_2 \leq \frac{1}{2}\log\left( 1 + \frac{(1-\theta)P}{\theta P +
\sw}\right).
\end{align*}
Furthermore, \eqref{eq:sbc:R1} and \eqref{eq:sbc:epi} will imply,
\begin{align*}
R_1 \leq \frac{1}{2}\log\left(1+\frac{\theta P}{\ss}\right).
\end{align*}
Thus, we have shown that $(R_1,R_2)\in \mathcal C_{\sbc}^{\nofb}$, and the
proof is complete.
\end{IEEEproof}
Theorem~\ref{thm:fb:gbc} can also be extended to more generalized feedback
settings with additive Gaussian  noise where the feedback noise may not be
independent of the forward channel noises. This extension will be
considered in a longer version of this paper.

\section{Vector Broadcast Channel with Feedback} \label{sec:mult}
It may appear that the additive Gaussian noise in the feedback plays a
critical role in the negative result that we presented in the last section.
In this section we study a vector GBC where the signal from one of the
antennas will be fed back {\em perfectly}, but this still does not enlarge
the capacity region in certain settings as we will describe below.

Let us consider the two-user memoryless broadcast channel shown in
Figure~\ref{fig:bc:1} below, where user~$1$ makes two independent
observations $(Y_{11},Y_{12})$ of each transmitted symbol. User~$2$
observes a scalar output $Y_2$. For simplicity, we will assume that the
noises on different antennas are independent.  Specifically, let $Z =
(Z_{2}, Z_{11}, Z_{12})$ be a zero mean Gaussian random vector with a
diagonal covariance matrix 
$$
K_z = \begin{bmatrix} \sw & 0 & 0 \\ 0 & \sas & 0 \\ 0 & 0 & \sbs \end{bmatrix}.
$$
Let us assume that the output symbols $Y_{12}$ are perfectly fed back to
the transmitter (causally). Notice that $Y_{11}$ as well as $Y_2$ are not
fed back. Let $\mathcal C_{\vbc}^{\pfb}$ and $\mathcal C_{\vbc}^{\nofb}$
denote the feedback and no-feedback capacity regions.
\begin{figure}[htbp]
\begin{center}
\scalebox{0.88}{
\begin{tikzpicture}[line width=0.5pt, >=triangle 45, text centered]
\coordinate (cenc) at (-0.5,0);
\coordinate  (cdec2) at  (4.75,1.25);
\coordinate  (cdec1) at  (4.75,-1.5);
\coordinate  (cz2) at  (2.5,1.25);
\coordinate  (cz1) at  (2.5,-0.5);
\coordinate  (cz12) at (2.5,-2.5);

\node (enc) at (cenc)[rectangle, draw, text width=1.6cm, minimum height=1cm]{Sender};
\node (z2)  at (cz2)[circle, draw] {$+$};
\node (z1)  at (cz1)[circle, draw] {$+$};
\node (z12)  at (cz12)[circle, draw] {$+$};
\node (dec2) at (cdec2)[rectangle, draw, text width=1.6cm, minimum height=1.0cm]{Receiver~2};
\node (dec1) at (cdec1)[rectangle, draw, text width=1.6cm, minimum height=3.0cm]{Receiver~1};
\draw[->] (z2) -- node[above, pos=0.4]{$Y_2$} (dec2); 
\draw[->] (z1) -- node[above, pos=0.4]{$Y_{11}$} (z1 -| dec1.west);
\draw[->] (z12) -- node (cfb) [pos=0.4]{} (z12 -| dec1.west);
\node at (cfb)[above]{$Y_{12}$}; 
\draw[<-] (enc) -- node[above, pos=0.6]{$W_1, W_2$} ++(-2.2,0) ;
\draw[->] (enc) -- node[above, pos=0.5]{$X$} ++(1.5,0) -- (z2);
\draw[->] (enc)++(1.5,0) -- (z1);
\draw[->] (enc)++(1.5,0) -- (z12);
\draw[->] (cfb.center) --++(0,-1.35)  -| (enc);
\draw[->] (dec2) -- node[above, pos=0.4]{$\hat W_2$} ++(2,0);
\draw[->] (dec1) -- node[above, pos=0.4]{$\hat W_1$} ++(2,0);
\draw[<-] (z2) --++(0,-0.75)node[below]{$Z_2$};
\draw[<-] (z1) --++(0,-0.75)node[below]{$Z_{11}$};
\draw[<-] (z12) --++(0,-0.75)node[below]{$Z_{12}$};
\end{tikzpicture}
}
\caption{A vector Gaussian broadcast channel with feedback\label{fig:bc:1}}
\end{center}
\end{figure}
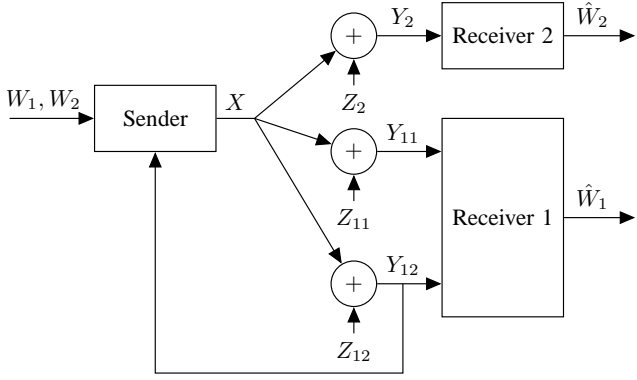

We study the case where  $\sas \leq \sw$, i.e., the second receiver has a
larger noise variance than the noise variance at the antenna of the first
receiver that is not fed back.  Under this assumption, we will show that
perfect causal feedback of $Y_{12}$ does not enlarge the capacity region.
Notice that the `no enlargement' result holds good irrespective of the link
quality of $Y_{12}$. This is surprising, since we are feeding back a major
portion of the output to receiver~1, particularly so when $\sbs <
\sas$.  We now state our main theorem.

\begin{theorem} \label{thm:two}
$ \mathcal C_{\vbc}^{\pfb} = \mathcal C_{\vbc}^{\nofb} $
if $\sas \leq \sw$.
\end{theorem}

The proof is along the same lines as that of Theorem~\ref{thm:fb:gbc}
presented in the last section.
\begin{IEEEproof}
Recall that the boundary of the no-feedback capacity region $\mathcal C_{\vbc}^{\nofb}$ is
given by
\begin{align*}
R_1 &\leq  \frac 12 \log\left(1 + \theta P\left[\frac 1{\sas} + \frac
1{\sbs} \right] \right),\\
R_2 &\leq  \frac 12 \log\left(1+\frac{(1-\theta)P}{\theta P + \sw}\right),
\end{align*}
for $\theta\in[0,1]$. This is just the capacity region of the scalar broadcast channel
resulting from receiver~1 pre-processing $Y_{11}$ and $Y_{12}$ by passing
it through a maximal ratio combiner.

By Fano's inequality,
\begin{align}
nR_2 &\approx I(W_2;Y_2^n)\\
        &= h(Y_2^n) - h(Y_2^n|W_2) \notag \\
	&\leq  \frac n2 \log \pi e (P+ \sw) - h(Y_2^n|W_2).
\end{align}
We know that there exists some $\theta \in [0,1]$ such that
\begin{align} \label{eq:r1:2}
h(Y_2^n|W_2) = \frac n2 \log 2 \pi e (\theta P + \sw).
\end{align}
Thus, 
\begin{align}
R_2 \leq \frac 12\log\left(1+\frac{(1-\theta)P}{\theta P + \sw}\right).
\end{align}
Let us now consider the rates for receiver~$1$. We will write
$(Y_{11},Y_{12})$ as ${\bf Y}_1$, and let ${\bf Y}_1^n$ denote the $2\times
n$ matrix of received values at receiver~$1$.  Again, by Fano's inequality
\begin{align}
nR_1 &\approx I(W_1;{\bf Y}_1^n|W_2) \\
	&= \sum_{i=1}^n  I(W_1; {\bf Y}_{1,i} |W_2, {\bf Y}_1^{i-1}).
\end{align}
Let $\frac 1{\sigma_e^2} = \frac 1{\sas} + \frac 1{\sbs}$ and consider the
following invertible transformation of ${\bf Y}_1$.
\[ \begin{bmatrix} Y_e \\ Y_e^{\perp} \end{bmatrix} = \sigma_e^2 \begin{bmatrix} \frac 1{\sas} &  \frac 1{\sbs} \\ -1 & +1 \end{bmatrix} {\bf Y} = \begin{bmatrix} X+Z_e \\ Z_e^{\perp} \end{bmatrix},\]
where $Z_e=\sigma_e^2 \left( \frac {Z_{11}}{\sas} + \frac {Z_{12}}{\sbs}
\right)$ is independent of $Z_e^{\perp}$. Note that $Z_e \sim \mathcal
N(0,\se)$ is an independent identically distributed (i.i.d.) process with
$\eE [Z_e Z_{12}^*] = \frac{\sas\sbs}{\sas+\sbs}$. Since the $Z_e^{\perp}$
process is i.i.d. and independent of the $Z_e$ process, we can  write,
\begin{align}
\sum_{i=1}^n  I(W_1 &; {\bf Y}_{1,i} |W_2, {\bf Y}_{1}^{i-1})\notag\\
 &= \sum_{i=1}^n h(Y_{e,i}|W_2, {\bf Y}_1^{i-1}) - h(Z_{e,i}) \notag \\
	&\stackrel{\text{(a)}}{\leq} \sum_{i=1}^n h(Y_{e,i}|W_2,Y_{11}^{i-1}) - h(Z_{e,i})\notag \\
	&= \sum_{i=1}^n h(X+Z_{e,i}|W_2,Y_{11}^{i-1}) - h(Z_{e,i}),\notag 
\end{align}
where in step (a) we removed $Y_{12}^{i-1}$ from the conditioning to obtain
the inequality. Since $Y_2$ has a larger noise variance than $Y_{11}$
(i.e., $\sw \geq \sas$) and both of them are not fed back to the
transmitter, we may write (also see footnote~\ref{fn:fb:sbc})
\begin{align}
n R_1 \leq \sum_{i=1}^n h(X_i+Z_{e,i}|Y_2^{i-1},W_2) - h(Z_{e,i}).
\end{align}
Consider an i.i.d. process $Z_d \sim \mathcal N(0,\sw-\se)$, independent of all other
processes mentioned before. Using El~Gamal's version of
EPI~\cite[Lemma~1]{Gamal81b}, 
\begin{align*}
2^{\frac 2n \sum_{i=1}^n h(Y_{e,i} |Y_2^{i-1},W_2)} &\leq  2^{\frac 2n h(Y_e^n + Z_d^n|W_2)} -  2 \pi e (\sw - \se) \\
		&= 2^{\frac 2n h(Y_2^n|W_2)} -  2 \pi e (\sw-\se)  \\
		&=  2 \pi e (\theta P + \sw) -  2 \pi e (\sw - \se) \\
		&= 2 \pi e (\theta P + \se). 
\end{align*}
Collecting all these together, we have
\begin{align}
R_1 \leq \frac 12 \log\left(1 + \frac{\theta P}{\se}\right).
\end{align}
This completes our converse, and we have shown whenever $\sas \leq \sw$,
perfect causal feedback from the antenna $Y_{12}$ does not change the
capacity region of  our broadcast channel. 
\end{IEEEproof}
While we have assumed independent noise processes at the antennas, this can
be extended to the case where 
the noises may be correlated. In fact, Theorem~\ref{thm:fb:gbc} can be
obtained as a corollary of such a generalization.

\section{Conclusion} \label{sec:conc}
We presented a class of two-user scalar Gaussian broadcast channels with
passive noisy feedback from one of the receivers for which feedback does
not enlarge the capacity region. Our result is in the form of a threshold
on the feedback noise variance above which feedback cannot achieve points
outside the no-feedback capacity region. We also saw a class of two-user vector
Gaussian broadcast channels where perfect feedback of some components of
the received signal from a user does not lead to an enlarged capacity
region.

Our study raises the question whether the threshold in
Theorem~\ref{thm:fb:gbc} is tight, i.e., is there a coding scheme which
can achieve points outside the no-feedback capacity region whenever the
feedback noise variance is below our threshold (and also in all cases where
Theorem~\ref{thm:fb:gbc} does not apply).

It must be emphasized that we only considered passive feedback. With active
feedback schemes no such thresholds may exist, e.g., one simple way to get
around the limitation is to conserve power on the feedback link and use it
only every so often (or equivalently, to send only a subsampled version of
the received signal, but repeat each sample several times) so that the
effective noise variance on the feedback link is below our threshold. This
suggests that it is unlikely that there are any such thresholds for active
feedback schemes with a noisy link. The same comment applies to
rate-limited feedback~\cite{Wigger14}.

\section*{Acknowledgements} 

This research was supported in part by Information Technology Research Academy (ITRA), Government of India under ITRA-Mobile grant ITRA/15(64)/Mobile/USEAADWN/01. 
SRBP's research was also supported by the Department of Science \& Technology (DST) India grant SB/S3/EECE/077/2013.
VP's research was partially supported by a Ramanujan Fellowship from the Department of Science \& Technology, Government of India.

\usetikzlibrary{arrows}

\appendix
\begin{IEEEproof}[Proof of Lemma~\ref{lem:epi}]

We prove this by induction on $n$. For $n=1$, the inequality follows from
entropy power inequality~\cite[pg. 22]{GamalKim11} since we may write
\[ h(Y_{2,1}|W_2) = h(Y_{1,1}+Z_{2,1}'|W_2),\]
where $Z_{2,1}'$ is zero-mean Gaussian with \[\Var(Z_{2,1}')=\sw-\ss\] and
independent of everything else. This simply follows from the fact that
$(Z_{1,1},Z_{2,1})$ is independent of $(W_2,X_1)$, and $\Var(Z_{1,1})=\ss
\leq \sw = \Var(Z_{2,1})$.

Suppose \eqref{eq:sbc:epi} is true for $n=m-1$. We may write
\[ h(Y_{2m}|(W_2,Y_2^{m-1})) = h(Y_{1m}+Z_{2m}'|(W_2,Y_2^{m-1})),\]
where $Z_{2m}'$ is zero-mean Gaussian with \[\Var(Z_{2m}')=\sw-\ss\] and
independent of everything else. This follows from the fact that
$(Z_{1m},Z_{2m})$ is independent of $(W_2,Y_2^{m-1},X_m)$ and
$\Var(Z_{1m})\leq \Var(Z_{2m})$.
By (conditional) entropy power inequality~\cite[pg. 22]{GamalKim11},
\[ 2^{2h(Y_{2m}|W_2,Y_2^{m-1})} \geq 2^{2 h(Y_{1m}|W_2,Y_2^{m-1})} +
2^{2 h(Z_2')}.\]
i.e.,
\[ 2h(Y_{2m}|W_2,Y_2^{m-1}) \geq \log\left( 2^{2 h(Y_{1m}|W_2,Y_2^{m-1})} +
2\pi e (\sw-\ss)\right).\]
So,
\begin{align*}
&\frac{2}{m}h(Y_2^m|W_2)\\
&= \frac{m-1}{m}\frac{2}{m-1}h(Y_2^{m-1}|W_2) + 
\frac{2}{m}h(Y_{2m}|W_2,Y_2^{m-1})\\
&\stackrel{\text{(a)}}{\geq}
\frac{m-1}{m}\log\left( 
    2^{\frac{2}{m-1} \sum_{i=1}^{m-1} h(Y_{1i}|W_2,Y_2^{i-1})} 
    + 2\pi e (\sw-\ss) \right)\\
&\qquad+
\frac{1}{m}\log\left(
    2^{2 h(Y_{1m}|W_2,Y_2^{m-1})} + 2\pi e (\sw-\ss)
\right)\\
&\stackrel{\text{(b)}}{\geq}
\log\left(
2^{\frac{2}{m}\sum_{i=1}^m
h(Y_{1i}|W_2,Y_2^{i-1})} + 2\pi e (\sw-\ss)
\right).
\end{align*}
where (a) follows from the induction hypothesis and the EPI above, and (b)
follows from convexity of $\log(2^u+v)$ in $u$ for $v\geq 0$.
\end{IEEEproof}


\begin{thebibliography}{10}
\providecommand{\url}[1]{#1}
\csname url@samestyle\endcsname
\providecommand{\newblock}{\relax}
\providecommand{\bibinfo}[2]{#2}
\providecommand{\BIBentrySTDinterwordspacing}{\spaceskip=0pt\relax}
\providecommand{\BIBentryALTinterwordstretchfactor}{4}
\providecommand{\BIBentryALTinterwordspacing}{\spaceskip=\fontdimen2\font plus
\BIBentryALTinterwordstretchfactor\fontdimen3\font minus
  \fontdimen4\font\relax}
\providecommand{\BIBforeignlanguage}[2]{{%
\expandafter\ifx\csname l@#1\endcsname\relax
\typeout{** WARNING: IEEEtran.bst: No hyphenation pattern has been}%
\typeout{** loaded for the language `#1'. Using the pattern for}%
\typeout{** the default language instead.}%
\else
\language=\csname l@#1\endcsname
\fi
#2}}
\providecommand{\BIBdecl}{\relax}
\BIBdecl

\bibitem{Shannon56}
C.~E. Shannon, ``The zero-error capacity of a noisy channel,''
  \emph{IRE Transactions on Information Theory}, vol.~19, 1956.

\bibitem{GamalKim11}
A.~El~Gamal and Y.-H. Kim, \emph{Network Information Theory}.\hskip 1em plus
  0.5em minus 0.4em\relax Cambridge University Press, 2011.

\bibitem{Gamal78}
A.~El~Gamal, ``The feedback capacity of degraded broadcast channels,''
  \emph{IEEE Transactions on Information Theory}, vol. IT-24, pp. 379--381, 1978.

\bibitem{Dueck80}
G.~Dueck, ``Partial feedback for two-way and broadcast channels,''
  \emph{Information and Control}, vol.~46, no.~1, pp. 1--15, 1980.

\bibitem{Gamal81b}
A.~El~Gamal, ``The capacity of the physically degraded Gaussian broadcast
  channel with feedback (corresp.),'' \emph{IEEE Transactions on Information Theory}, vol.~27, no.~4, pp. 508--511, Jul 1981.

\bibitem{Ozarow84b}
L.~H. Ozarow and S.~K. Leung-Yan-Cheong, ``An achievable region and outer bound
  for the Gaussian broadcast channel with feedback,'' \emph{IEEE Transactions on Information Theory}, vol. IT-30, pp. 667--671, 1984.

\bibitem{Sibi08}
S.~R. Bhaskaran, ``Gaussian {BC} with feedback,'' \emph{IEEE Transactions on Information Theory}, pp. 5252--5257, Aug 2008.

\bibitem{Shayewitz13}
O.~Shayevitz and M.~Wigger, ``On the capacity of the discrete memoryless
  broadcast channel with feedback,'' \emph{IEEE Transactions on Information Theory}, vol.~59, no.~3, pp. 1329--1345, Mar 2013.

\bibitem{Pradhan13}
R.~Venkataramanan and S.~Pradhan, ``An achievable rate region for the broadcast
  channel with feedback,'' \emph{IEEE Transactions on Information Theory},
  vol.~59, no.~10, pp. 6175--6191, Oct 2013.

\bibitem{Wigger13}
Y.~Wu and M.~Wigger, ``Any positive feedback rate increases the capacity of
  strictly less-noisy broadcast channels,'' in Proc. of Information Theory
  Workshop (ITW) 2013, pp. 1--5, Sept 2013.

\bibitem{Wigger14}
------, ``Coding schemes for discrete memoryless broadcast channels with
  rate-limited feedback,'' \emph{CoRR}, vol. abs/1401.6219, 2014.

\bibitem{LapWig10}
A.~Lapidoth and M.~Wigger, ``On the AWGN MAC with imperfect feedback,''
  \emph{IEEE Transactions on Information Theory}, vol.~56, no.~11, pp.
  5432--5476, Nov 2010.

\bibitem{AmStWi14}
B.~Amor, Y.~Steinberg, and M.~Wigger, ``MIMO MAC-BC duality with
  linear-feedback coding schemes,'' \emph{CoRR}, vol. abs/1404.2584, 2014.

\bibitem{CovTho91}
T.~M. Cover and J.~A. Thomas, \emph{Elements of Information Theory}.\hskip 1em
  plus 0.5em minus 0.4em\relax Wiley, 1991.

\bibitem{Blachman65}
N.~M. Blachman, ``The convolution inequality for entropy powers,''
  \emph{IEEE Transactions on Information Theory}, vol.~11, no.~2, pp.
  267--271, Apr 1965.

\end{thebibliography}
\end{document}